\documentstyle[aps,epsf,epsfig,12pt,prl]{revtex} 

\begin{document}
\bibliographystyle{revtex4}

\title{Autonomous Bursting in a Homoclinic System}

\author{R. Meucci, A. Di Garbo$^{*}$, E. Allaria 
and F.T. Arecchi$^{+}$}
\address{Istituto Nazionale di Ottica Applicata, Largo E. Fermi 6, 50125
Florence, Italy \\
$^{*}$ also Istituto di Biofisica CNR, Pisa, Italy \\
$^{+}$ also Department of Physics, University of Firenze, Italy}

\date{\today}

\maketitle 

\hfill\break

\begin{abstract}
A continuous train of irregularly spaced spikes, peculiar of homoclinic
chaos, transforms into clusters of regularly spaced spikes, with quiescent 
periods in between (bursting regime), by feeding back a low frequency 
portion of the dynamical output. Such autonomous bursting
results to be extremely robust against noise; we 
provide experimental evidence of it 
in a CO$_2$ laser
with feedback. The phenomen here presented display qualitative
analogies with bursting phenomena in neurons.

PACS\ numbers : 
05.45.-a; 89.20.-a; 89.75.-k.
\end{abstract}

\hfill\break

Spiking occurs in many physical and biological systems
\cite{uno}; in neurons it is associated with the generation of action
potentials \cite{due}. In general, spikes occur irregularly and the
precise time position of each spike within a train is considered to code
relevant information \cite{tre}. A single neuron can also show a more complicated
firing pattern, that is, bursting. The term bursting refers to short trains of
rapid spike oscillations intercalated by quiescent intervals, which periodically repeat.
This behavior is found in many excitable
biological systems as well as in chemical reactions \cite{quattro,cinque,sei}. In the neural
context bursting phenomena have been found in different cell types: examples
include thalamic neurons \cite{sette}, hippocampal pyramidal neurons \cite{otto} and
pancreatic $\beta$-cells \cite{nove,dieci}. The understanding of mechanisms underlying
bursting is based on the fundamental work of Rinzel \cite{dodici}, accordingly to whom the
neuron dynamics implies two different time scales and it can be generally
represented in the following form

\begin{equation}
\begin{tabular}{c}
$\dot{X}=F(X,Y)$ \\ 
$\dot{Y}=G(X,Y)$
\end{tabular}
\end{equation}
Here, the vector $X$ describes the fast variables generating spikes and 
$Y$ the slow ones contributing a modulation. The fast variables are
associated with the membrane potential, Na$^{+}$ and K$^{+}$ conductance
and other variables evolving on time scales comparable with the duration of
the spike, whereas
the slow variables refer to the concentration and conductance
of Ca$^{2+}$ ions \cite{tredici}.
Bursting arises as the evolution of the slow variables
switches the fast dynamics between steady state and oscillatory dynamics.
Thus a feedback process from $Y$
confers to $X$ the feature of a bursting pattern
consisting of a fast spiking regime riding on a 
slow modulation \cite{tredici,quattordici}.
On the other hand, spiking phenomena are observed and easily modeled in
other physical areas. Spike emission occurs in multi-mode class B lasers
(solid state and semiconductor) operating close to threshold.
This type of emission does not occur in a single-mode class B laser as CO$_2$
, but it can be induced by a suitable feedback \cite{quindici}.

Here we present evidence of
the autonomous passage from a continuos train of irregular spikes to periodic bursts
of regularly spaced spikes in a laser system. Precisely, we take as
the fast system a laser undergoing homoclinic chaos of the Shilnikov type
\cite{quindici}. In this system, the presence of {\sl non-autonomous} bursting 
induced by an external modulation has, already been
reported \cite{sedici}, with the irregular timing of the homoclinic spikes being
regularized into a synchronized sequence of pulses. We consider such a behavior
as generic for all spiking systems based on an
activator-inhibitor competition (as e.g. biological clocks). Indeed the
close approach to a saddle point, peculiar of homoclinic chaos, provides a
local slowing down of the few variables (3-dimensional subset of $X$) describing the
inflow to, and the outflow from, that unstable region. This means that all
the other variables (complementary subset of $X$) do not play a crucial role
around the singularity, whence such a generic homoclinic behavior applies to
many different models which have in common the presence of a saddle point.
The core features around the saddle point require three coupled variables;
however, in order to recover the global aspects of phase-space orbits away from the
saddle point, we will use a 6-dimensional model as reported later in
Eqs.(2). Motivated by the fact that
bursting in biological systems {\sl does not} require
an {\sl external forcing}, we explore the feasibility of
{\sl autonomous} bursting in the laser. This is achieved by a
secondary feedback loop, where a fraction of the output is fed back to a
loss modulator after passage through a low pass filter.
The slow dynamics thus provides the mechanism for
bursting the laser intensity.

The experiment has been performed on a single mode CO$_2$ laser with an 
intracavity loss modulator; the detected output
intensity is fed back in order to control the cavity losses (Fig. 1).
The DC level and amplitude of the feedback signal are controlled by
adjusting the bias (B$_0$) and the gain (R) of the amplifier. We set these
two parameters so that the laser intensity displays a regime of
Shilnikov chaos (the saddle point being a saddle focus \cite{diciasettesh}).
The laser intensity, after a time interval where it
practically approaches the zero intensity, displays a large spike followed
by a damped train of fast oscillations and a successive train of growing
oscillations (Fig. 2). Damped and growing trains represent respectively
the approach to, and the escape from, the saddle
focus $S$ from where the trajectory rapidly returns to zero and then starts a
new orbit. In the chaotic region, the escape time from the saddle focus
has different duration from pulse to pulse.
The power spectrum shows a broadened peak at the average repetition
frequency of the pulses, $\nu _n$ $\simeq $ 2 kHz. 
The non vanishing frequency content below $\nu
_n$ suggests that it is possible to enhance one of those frequencies by
feeding back a filtered fraction of the output. 
Thus, we filter the detected signal by 
a low pass filter with a cut off frequency ($
\nu _c$) lower than $\nu _n$ and then re-inject
it through a secondary feedback loop.
As a result, taking also into account the phase response of the filter, we
select a bursting frequency somewhat lower than, but proportional to, $\nu _c$.
The other low ($\nu
<\nu _c$) frequency components are efficiently suppressed. 
This way, we obtain clusters of laser pulses periodically spaced
at the bursting frequency.
We have used ''Rockland mod. 452'' filters with a 
slope of 40 dB/decade.
It was experimentally tested that a single filter is not sufficient;
we were obliged to cascade two filters. In fact, we will see later that the 
numerical model requires at least three poles.
The secondary feedback loop contains
also a variable gain amplifier, AC coupled in order not to alter the primary
feedback loop setting. 
The filter yields bursting sequences with adjustable periods, reset by changing 
$\nu _c$ (Fig. 3).

The reported experimental evidence on a single laser system should not mask the cooperative
character of the phenomenon here investigated.
This character is demonstrated by the fact that such
a feedback is dynamically equivalent to the coupling of two
independent systems. For this purpose, we have measured the correlation time of the chaotic
signal, which results to be around 100 $\mu s$, consistently lower than the average interspike interval
($\nu _n^{-1}\simeq 500 \mu s$). Once we select a feedback frequency around 500 Hz, well below
$\nu _n$, we are assured that the feedback onset time ($\simeq 2 ms$)
is much longer than the decorrelation time.
Hence, our feedback is equivalent to the cooperative interaction of independent lasers.

The dynamics is modeled by coupling the
laser dynamics, described by a six-dimensional state vector $X$ \{$x_i : i=1-6$\} \cite{diciasettepis},
to the slow dynamics of a three-dimensional filter $Y$ \{$y_j : j=1-3$\} : 
\begin{equation}
\begin{array}{l}
\dot{x}_1=k_0x_1(x_2-1-k_1\sin ^2(x_6)) \\ 
\dot{x}_2=-\gamma _1x_2-2k_0x_2x_1+g x_3+x_4+p_0 \\ 
\dot{x}_3=-\gamma _1x_3+x_5+gx_2+p_0 \\ 
\dot{x}_4=-\gamma _2x_4+g x_5+zx_2+zp_0 \\ 
\dot{x}_5=-\gamma _2x_5+zx_3+g x_4+zp_0 \\ 
\dot{x}_6=-\beta (x_6+ B_0 - R(x_1-\phi
(y_3-\bar{y}_3))) \\ 
\dot{y}_1=-\theta \left( y_1-x_1\right)  \\ 
\dot{y}_2=-\theta \left( y_2-y_1\right)  \\ 
\dot{y}_3=-\theta \left( y_3-y_2\right) .
\end{array}
\end{equation}
In these equations the variable $x_1$ represents the normalized laser
intensity, $x_2$ and $x_3$ are respectively proportional to
the population difference and sum of the two molecular levels resonant with
the field, $x_4$ and $x_5$ represent the difference and sum of the
populations of the rotational manifolds, which supply excitation energy to
the resonant transition. The variable $x_6$ is proportional to the feedback
voltage and it affects the cavity loss parameter via the term $k_1\sin
^2(x_6)$. The time is rescaled according to $t=t\cdot 7\cdot 10^5$. $\gamma
_1$ ,$\gamma _2$ , $g$ and $\beta $ are decay rates, $p_0$ is the pump
parameter, $z$ is the number of sublevels contained in the rotational
manifolds. In the $x_6$ equation $B_0$ and $R$ are
respectively the bias and gain of the primary feedback amplifier and $\phi $
is the gain of the secondary feedback amplifier. 
The slow dynamics corresponds to a low pass filter with a third order pole
at $\theta =2\pi \nu _c$ , therefore it is modeled by the cascade of three
linear integrators. The numerical values of the parameters have been 
chosen so that the free running behavior of the
laser displays Shilnikov chaos \cite{diciasettepis}(Table I).

Fig. 4 shows the model behavior at the onset of the
stable bursting corresponding to a perturbation of 2\%. It is important to
note that the use of a single pole filter does not provide bursting because
the frequency component corresponding to the average repetition frequency of
the chaotic pulses ($\nu _n\simeq 2$ kHz) is not sufficiently rejected. In
this case, the action of the secondary feedback results in a modulation at
the average repetition frequency. 

According to the general bursting model of Eq.(1), the filter variables
represent the slow subsystem $Y$. In particular, the $Y$ equations are similar
to those used in the Hindmarsh and Rose model for neuronal bursting \cite{diciotto}.

In our case the bursting phenomenon is related to 
homoclinic chaos, which requires a fine parameter tuning in order to set
the orbits around the saddle focus \cite{sedici,diciasettesh,diciasettepis}.
As the control parameter B$_0$ is moved above or below the homoclinic value,
the dynamical system enters respectively a periodic behavior or it goes
to a meta-stable fixed point (excitable system). This is clearly shown by a 
$\pm $ 1\% stepwise change of B$_0$ as shown in Fig. 5-a,b. 
Furthemore, in the case of a positive perturbation, the repetition frequency $\nu _r$
of the periodic spike train is a monotonic 
function of the perturbation amplitude, up to 5\%
(Fig. 5-c).

Such an evidence demonstrates a fact that may be 
relevant for neurodynamics, that is, {\sl noise is not necessary} 
to induce a bursting phenomenon. We have checked the independence
of autonomous bursting from noise by applying to the laser pump an
additive white noise up to 10 \% of the pump's nominal value, without observing 
modifications. This demonstrates that autonomous bursting does not
belong to the class of noise induced phenomena as e.g. stochastic resonance
\cite{gamma} or coherence resonance \cite{piko}.

To make comparison with neurophysiological behavior,
some real neurons do
not show bursting when the synaptic connections with other cells are
blocked. In such a condition, a single neuron typically shows chaotic
spiking activity. Examples are provided by pancreatic $\beta $-cells \cite{dicianove} or
by neurons of the pyloric Central Pattern Generator in the stomatogastric
ganglion of crustaceans \cite{venti}. In these cases the appearance of bursting is
related to the synaptic interaction among the cells \cite{ventuno,ventidue,ventitre}. 
This behavior can be interpreted in the framework of our model if we take the secondary
feedback term as inducing a cooperative coupling among identical coupled lasers.
As for the frequency $\nu _r$ 
of the regular spiking within
each burst, Fig. 5-c shows that it depends sensitively
upon the B$_0$ perturbation.
This suggests that an accurate control of Calcium concentration 
might be the neurodynamic tool for a reliable and sensitive coding of different stimuli,
without need to recur to noise \cite{tre,venticinque}.

In conclusion, the chaotic spiking regime peculiar of homoclinic chaos
has been converted to an {\sl autonomous bursting} regime by means of a feedback
loop acting on a long temporal scale as compared to the average repetition time
of the chaotic pulses. The observed regimes display characteristics similar
to those observed in neurons \cite{sette,otto,tredici} 
and are reproduced by a
model where the fast dynamics is modulated by a slow one through a 
secondary feedback loop; the frequency of the modulation sets the bursting
period, while its amplitude sets the spike repetition frequency within each burst.

\hfill\break
We acknowledge partial support from the European Contract No. HPRN-CT-2000-00158. 
A. D. G. is supported by the European Contract No. PSS 1043.


\newpage

\begin{center}
{\bf Figure captions}
\end{center}

\hfill\break

\begin{figure}[h]
\centerline{
\epsfig{figure=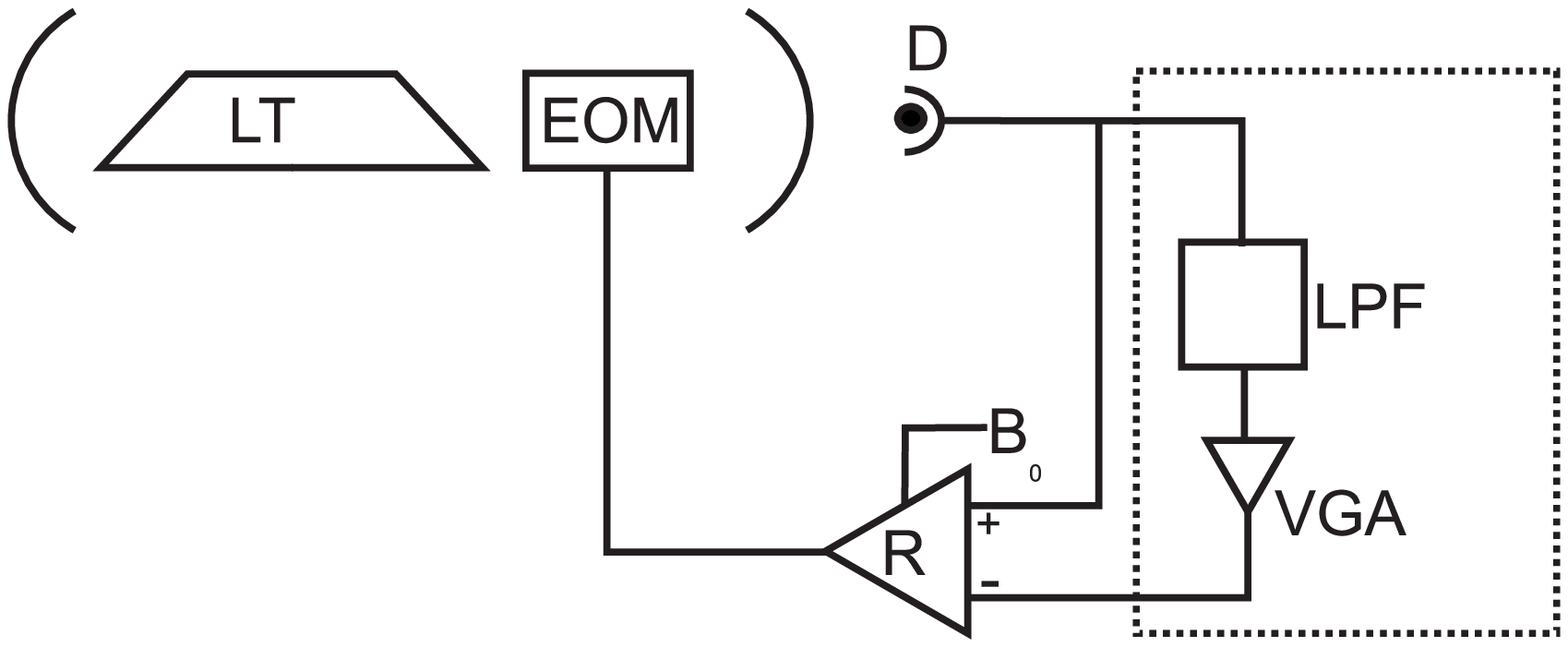,width=12cm}}
\hfill\break

\caption{Experimental setup consisting of a CO$_2$ laser with a primary feedback loop
imposing a regime of homoclinic chaos and a secondary slow feedback loop
(dashed block) which produces bursting. LT, laser tube; EOM, electro-optic
modulator; D, HgCdTe detector; LPF, low pass filter; VGA variable gain
amplifier (AC coupled); R and B$_0$ respectively gain and bias of the
amplifier in the primary feedback loop.}
\label{figura1}
\end{figure}

\hfill\break
\begin{figure}[h]
\centerline{
\epsfig{figure=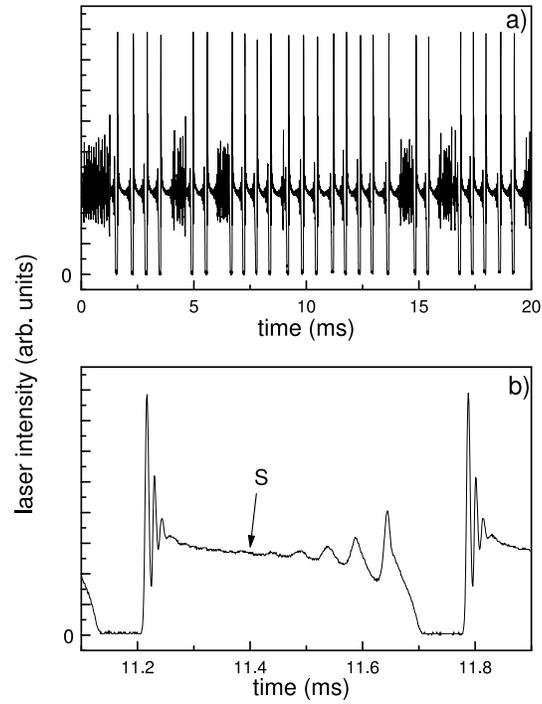,width=7cm}}
\hfill\break

\caption{a) Experimental time series of the laser output intensity in the homoclinic chaos
regime (zero intensity level indicated). b) Zoom of a single pulse showing the approximate
location of the saddle focus $S$ at the start of the growing oscillation; 
notice that the preceding large damped oscillation occurs away from $S$ and it is 
not part of the local dynamics at $S$.}
\label{figura2}
\end{figure}

\hfill\break
\begin{figure}[h] 
\centerline{ 
\epsfig{figure=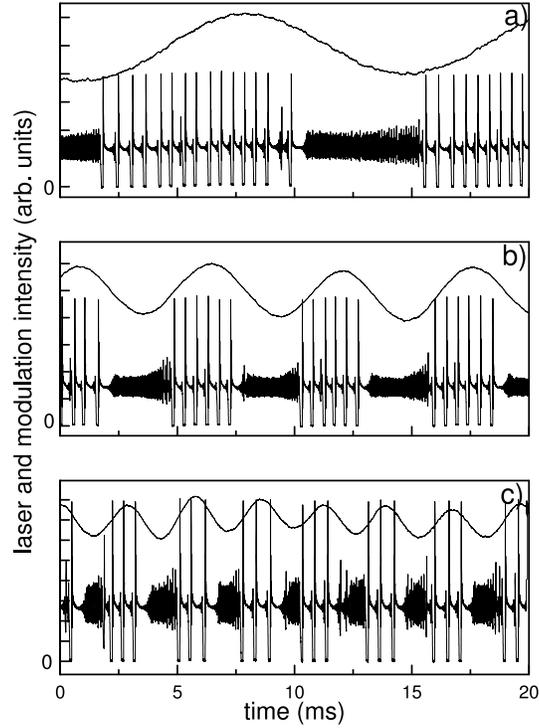,width=7cm}} 
\hfill\break 

\caption{Experimental time series of bursting obtained with the Rockland filter
for three different values of the cut off frequency a) $\nu_c=100$ Hz, b)$\nu 
_c=300$ Hz, c) $\nu _c=600$ Hz. The low frequency sinusoids are trains of the
filtered feedback signal reported on a different scale; in fact the feedback
amplitude is 2 \% of the spike amplitude. The effective bursting frequencies
are respectively a) 80 Hz, b) 200 Hz and c) 350 Hz, that is, proportional to,
even though lower than, the corresponding cutoff $\nu _c$.}
 
\label{figura3}
\end{figure}
 
\hfill\break
\begin{figure}[h]
\centerline{ 
\epsfig{figure=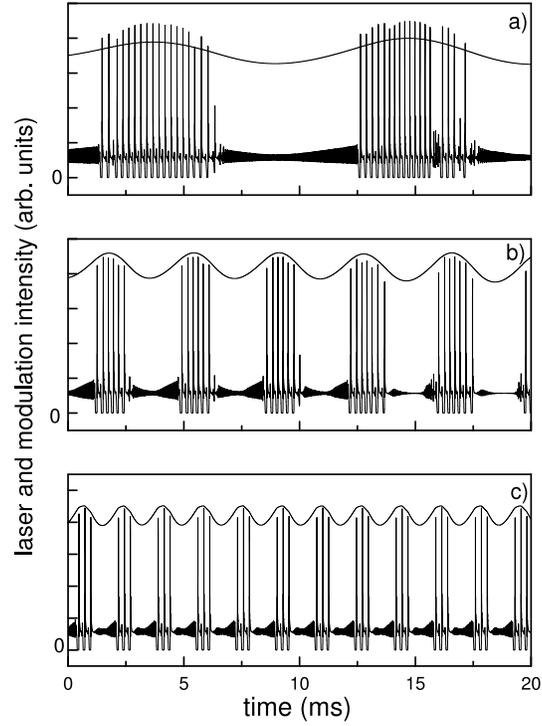,width=7cm}}
\hfill\break

\caption{
Numerical results corresponding to the three cases reported in Fig. 3. a) 
$\theta =4.5\cdot 10^{-4}$ ($\nu _c=100$ Hz), b) $\theta =1.3\cdot 10^{-3}$
($\nu _c=300$  Hz), c) $\theta =2.7\cdot 10^{-3}$ ($\nu _c=600$ Hz).
}
\label{figura4}
\end{figure}

\hfill\break
\begin{figure}[h]
\centerline{
\epsfig{figure=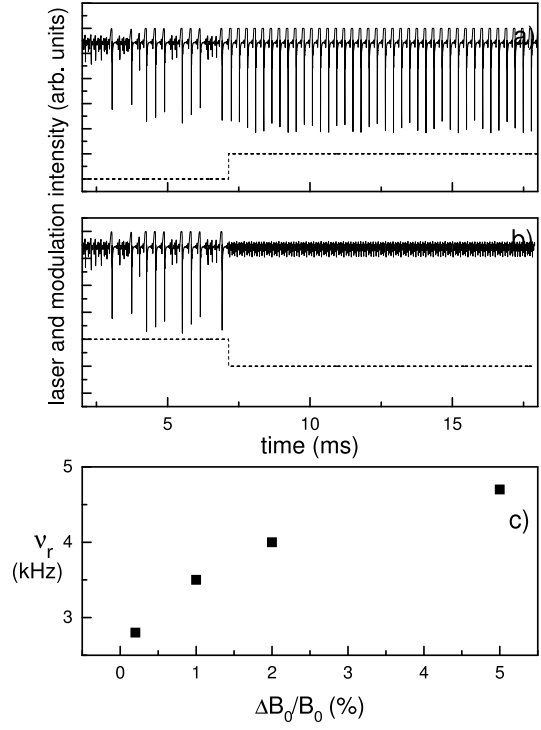,width=7cm}}
\hfill\break

\caption{Stepwise increase a)  and decrease b) of control parameter B$_0$ by 
$\pm $ 1\% (dashed line) brings the system from homoclinic behavior to respectively
periodic or excitable behavior.
c) In the case a) the frequency repetition 
$\nu _r$ of the spikes within the regular train increases monotonically up 
to a $\Delta B_0 / B_0 \sim$ 5 \% ; above this value, there is a saturation.
}
\label{figura5}
\end{figure}

\newpage
\hfill\break
\hfill\break

\begin{tabular}{|c|c||c|c|}
\hline
$\gamma _1$ & $10.0643$ & $k_0$ & $28.5714$ \\ \hline
$\gamma _2$ & $1.0643$ & $k_1$ & $4.556$ \\ \hline
$z$ & $10$ & $p_0$ & $0.016$ \\ \hline
$\beta $ & $0.4286$ & $R$ & $160$ \\ \hline
$g$ & $0.05$ & $B_0$ & $0.1031$ \\ \hline
$\theta $  &  $[4.5\cdot 10^{-4}\div 2.7\cdot 10^{-3}]$ &  $\phi $ & $4$
\\ \hline
\end{tabular}

\hfill\break

TableI 

Parameter values used in the simulations.

\end{document}